# The Geochemical Potential for Metabolic Processes on the Sub-Neptune Exoplanet K2-18b

Christopher R. Glein

Space Science Division, Space Sector, Southwest Research Institute, 6220 Culebra Road, San Antonio, TX 78238-5166, United States

Email: christopher.glein@swri.org

**Abstract**

Quantifying disequilibria is important to understand whether an environment could be habitable. It has been proposed that the exoplanet K2-18b has a hydrogen-rich atmosphere and a water ocean, making it a "hycean world". The James Webb Space Telescope recently made measurements of methane, $CO_2$, and possibly dimethyl sulfide (DMS) in the atmosphere of this planet. The initial interpretation of these data is that they may support the occurrence of hycean conditions. Here, I attempt to take a next step in exploring the prospects for habitability. I use constraints on the abundances of atmospheric gases to calculate how much chemical disequilibrium there could be, assuming K2-18b is a hycean world. I find that the presence of oxidized carbon species coexisting with abundant $H_2$ (1-1000 bar) at cool to warm (25-120°C) conditions creates a strong thermodynamic drive for methanogenesis. More than ~75 kJ (mol C)$^{-1}$ of free energy can be released from $CO_2$ hydrogenation. Partially oxidized carbon compounds such as DMS (if present) also have potential to provide metabolic energy, albeit in smaller quantities. Because of the thermodynamic instability of $CO_2$ under hycean conditions, other reductive reactions of $CO_2$ are likely to be favored, including the synthesis of amino acids. Glycine and alanine synthesis can be energy-releasing or at least much less costly on K2-18b than in Earth's ocean, even when $NH_3$ is scarce but not totally absent. These first bioenergetic calculations for a proposed ocean-bearing exoplanet lay new groundwork for assessing exoplanetary habitability.

# 1. Introduction

The exoplanet K2-18b (Montet et al. 2015) is of outstanding interest. This world orbits within the conventional habitable zone of its star (Kopparapu et al. 2013) and receives a stellar flux comparable to that of Earth (K2-18b has an equilibrium temperature of ~260 K for an Earth-like bond albedo). Hence, K2-18b could host a surface liquid water ocean. However, there are key differences between this planet and Earth; K2-18b is much larger (~2.6$R_\oplus$; Benneke et al. 2019) and more massive (~8.6$M_\oplus$; Cloutier et al. 2019) than Earth. Even with greater compression, it has a bulk density (~2700 kg m$^{-3}$; Benneke et al. 2019) that is lower than Earth's (5515 kg m$^{-3}$).

The internal structure of K2-18b cannot be uniquely determined from the density alone, although its low density is indicative of a volatile-rich composition. Based on modeling to match the density, Madhusudhan et al. (2020) found that K2-18b's structure is likely to be between the endmembers of a thick $H_2$-rich atmosphere overlying a rock-rich or rock + water interior, and a thin $H_2$-rich atmosphere overlying a deep hydrosphere on top of a smaller rocky core. The latter "ocean world" case is intriguing, and if it is correct, then K2-18b may comprise a new type of potentially habitable environment, referred to as a "hycean world" (Madhusudhan et al. 2021), where there could be abundant hydrogen gas in contact with an ocean. Earlier spectroscopic evidence for a low atmospheric mean molecular weight also suggested the existence of a $H_2$/He-dominated atmosphere (Benneke et al. 2019; Tsiaras et al. 2019).

Recent observations by the James Webb Space Telescope (JWST) advanced the suggestion that K2-18b might be a hycean world. Madhusudhan et al. (2023a) performed transmission spectroscopy of K2-18b from 0.9 to 5.2 μm and found that the spectrum can be reproduced by model atmospheres containing ~1% abundances (by volume) of $CH_4$ (0.17-7.1% at 1σ) and $CO_2$ (0.093-5% at 1σ), with a mixture of $H_2$ and He making up the balance. These abundances were found to be consistent with what would be expected from atmospheric chemical modeling if K2-18b is a hycean world (Hu et al. 2021; Madhusudhan et al. 2023b). However, a more recent photochemical model predicts that an uninhabited hycean world would have too little methane (Wogan et al. 2024). In this case, an unknown source of methane would need to be invoked to explain the observed abundance. This result may call for an inhabited hycean world (Madhusudhan et al. 2023a), or for the hycean hypothesis to be disfavored (Wogan et al. 2024). Alternatively, since a hypothetical hycean world would be utterly unfamiliar to us, we should not discount the possibility of unexpected chemistry occurring in the ocean or deeper interior (e.g., Miller et al. 2019; Sverjensky et al. 2020).

One of the most informative findings from JWST is what it did not detect. Neither $NH_3$ (<$10^{-4.46}$ at 2σ) nor CO (<$10^{-3.00}$ at 2σ) was detected (Madhusudhan et al. 2023a). This challenges "mini-Neptune" interpretations because $NH_3$ and CO are produced under hot, reducing conditions (Madhusudhan et al. 2023a). Several models have predicted this (Blain et al. 2021; Hu 2021; Tsai et al. 2021; Yu et al. 2021; Wogan et al. 2024). For example, the mini-Neptune model presented in the most recent study by Wogan et al. (2024) has $NH_3$ and CO mixing ratios of $10^{-3.15}$ (0.07%) and $10^{-2.5}$ (0.3%), respectively, in the main region of K2-18b's photosphere. These values are too high. This model also predicts a $CO/CO_2$ number ratio (~5) that seems inconsistent with the data ($CO/CO_2 < 1$; Madhusudhan et al. 2023a). Thus, the non-detection of $NH_3$ and CO may imply that scenarios having a thick $H_2$-rich atmosphere are not consistent with present compositional constraints. The exception would be if the observational data are more uncertain than what was found (Madhusudhan et al. 2023a). Another interesting idea that was recently proposed is that $NH_3$ could dissolve in a reduced silicate magma ocean (Shorttle et al. 2024), potentially explaining why $NH_3$ was not detected by JWST. However, this scenario may also be inconsistent with the lack of CO detection (Madhusudhan et al. 2023a).

If we consider that a best match to the data remains elusive, then the hycean interpretation may be one of many possibilities. Taking this idea further, we may then ask, would conditions in a hycean ocean be tolerable for life as we know it? No measurements have been made of temperatures in the lower atmosphere of K2-18b. Climate models predict that temperatures can reach >200-300°C in the deeper atmosphere, owing to $H_2$ collision-induced absorption that provides greenhouse heating and the possibility of inefficient convective heat transfer (Innes et al. 2023; Leconte et al. 2024). It has been suggested that high albedo aerosols could counteract this effect (Madhusudhan et al. 2023a). Also, if water is abundant and the lower atmosphere is as hot as predicted, then a supercritical water ocean may be present (Mousis et al. 2020). Such an ocean could introduce plentiful water vapor into the atmosphere, making its mean molecular weight too large (Scheucher et al. 2020; Pierrehumbert 2023). Water vapor was not detected by JWST (the upper limit on its mixing ratio is $10^{-3.06}$ at 2σ) at altitudes above ~100 mbar (Madhusudhan et al. 2023a).

The current body of knowledge conveys a sense that we are at an early stage of understanding. Accordingly, numerous possibilities should be investigated to cast a wide net that may capture what is happening on K2-18b. In the present paper, I adopt the hycean interpretation (Madhusudhan et al. 2023a) as it is clearly of great interest. The goal of this work is to see what else the JWST data can tell us about the habitability of K2-18b, assuming it is a hycean world. In addition, we might be interested in what we can learn about hycean worlds more broadly.

To be habitable, an environment must be able to support life, which apparently requires liquid water, essential chemical elements/organic building blocks, and a source of energy (e.g., McKay 2014). Here, I focus on the latter two requirements. If there is an ocean on K2-18b, how much chemical energy might be available for life to exploit? Can critical organic building blocks be produced? I will show how a theoretical framework can be constructed and observational constraints incorporated to make geochemical inferences about exoplanetary habitability. Quantification of disequilibrium using atmospheric composition was pioneered through applications to the evolution of Earth's atmosphere (Krissansen-Totton et al. 2018) and to the atmospheres of other solar system bodies (Krissansen-Totton et al. 2016). However, it was not possible to perform a similar evaluation for a potentially habitable exoplanet. We now have an opportunity. I will attempt to elucidate a pathway between the JWST data from K2-18b and the planet's energetic potential to support metabolic processes, so that we may be given some clues to the questions above. Note that the present study employs chemical thermodynamics and is concerned with free energy relationships between the present state of an ocean-atmosphere system on K2-18b and equilibrium states. I do not attempt to address rates of reactions, or how (bio)geochemical cycles might work on K2-18b. Nonetheless, quantifying disequilibria, as done here, represents an important step toward more comprehensive models that could be developed in the future to constrain power supplies from chemistry on an exoplanet.

## 2. Approach

### 2.1. Thermodynamic State of the Atmosphere

I assume that the atmospheric composition near the surface of an ocean on K2-18b can be approximated by the following summation

$$y_{H_2} + y_{He} + y_{CH_4} + y_{CO_2} + y_{H_2O} \approx 1, \tag{1}$$

where $y_i$ designates the mole fraction/mixing ratio of gas species $i$. Equation 1 can be recast in terms of parameters that are easier to determine, which yields

$$y_{H_2} + (He/H_2) y_{H_2} + y_{CH_4,dry} (1 - y_{H_2O}) + y_{CO_2,dry} (1 - y_{H_2O}) + y_{H_2O} \approx 1. \tag{2}$$

To solve this equation, I adopt a protosolar He/H$_2$ ratio of 0.20 (Lodders 2021), consistent with the idea that a sub-Neptune would have captured H$_2$ and He from the surrounding nebular gas during the formation of the planet (Bean et al. 2021). The mole fractions of $CH_4$ and $CO_2$ in equation 2 can be expressed relative to dry air (as in Madhusudhan et al. 2023a) via

$$y_{i,\text{dry}} = \frac{y_i}{y_{H_2} + y_{He} + y_{CH_4} + y_{CO_2}} = \frac{y_i}{1 - y_{H_2O}}. \tag{3}$$

I developed a method to estimate the water vapor content of the near-surface atmosphere based on phase equilibrium with the ocean. This requires the fugacity (*f*) of $H_2O$ that would be in equilibrium with an ocean. Fugacity can be thought of as a corrected partial pressure (Fegley 2013). If the ocean is relatively pure water, then the fugacity will only be a function of temperature (*T*) and total pressure (*P*). For comparison, Earth's seawater has a $H_2O$ fugacity that departs from that of pure water by only ~2% (Robinson 1954). I used the program Reference Fluid Thermodynamic and Transport Properties (REFPROP; Huber et al. 2022)—with the water equation of state of Wagner & Pruß (2002)—to understand how the fugacity of liquid water varies with *T* and *P*. REFPROP data showed linear relationships between $\log f_{H_2O(\text{liq})}$ and *P* at constant *T*, consistent with water being an incompressible fluid. Parameters for the best fits are provided in Table A1.

At a given *P-T*, I calculate the mole fraction of $H_2$ as a function of the mole fraction of water vapor using equation 2 with observational values for the dry mole fractions of $CH_4$ and $CO_2$ (Madhusudhan et al. 2023a). I then use REFPROP—with the mixing model of Kunz & Wagner (2012)—to compute the fugacities of all species (see Appendix B for a strategy for dealing with poorly characterized gas mixtures). The gas-phase composition with a $H_2O$ fugacity that matches the fugacity of liquid water at the temperature and pressure of interest is in equilibrium with liquid water. To guide the initial guess for $y_{H_2O}$, I assume $y_{H_2O} = p_{\text{sat},H_2O(\text{liq})} / P$, where $p_{\text{sat},H_2O(\text{liq})}$ stands for the saturation vapor pressure of liquid water. We can express the saturation vapor pressure of liquid water as a function of temperature using the following equation

$$\log p_{\text{sat},H_2O(\text{liq})}(\text{bar}) = 18.7 - 2855.3/T(\text{K}) - 4.294 \times \log T(\text{K}), \tag{4}$$

which was obtained by regressing numerical data from REFPROP. Those data are stated to have an uncertainty of 0.025% (Wagner & Pruß 2002). Equation 4 reproduces vapor pressures from REFPROP to within ~1% from 0 to 150°C. This regressed region includes the temperature range that is most relevant to habitability (see Section 3.1.1).

After calculating the major gas-phase composition, I perform additional non-ideal gas calculations for variable mole fractions of $NH_3$. $NH_3$ is unlikely to affect the fugacities of the major atmospheric constituents because it is not known to be abundant in the atmosphere of K2-18b (Madhusudhan et al. 2023a). On the other hand, even if $NH_3$ has a minor or trace atmospheric abundance, it is important to constrain its fugacity because it influences the stabilities of amino acids (see Section 3.2).

*2.2. Quantification of Redox Disequilibria*

2.2.1. Methanogenesis

Energy is a fundamental requirement of life (Hoehler et al. 2020). Now that some gases have been identified in the atmosphere of K2-18b, a natural question is whether any of them could contribute to energy supplies that might fuel life. Below, I focus on two slices of the chemotrophic landscape of K2-18b that can be linked to current observations of atmospheric composition. Photosynthesis is another possibility (Bains et al. 2014) that may be feasible if liquid water is present at temperatures below ~75°C (e.g., Cox et al. 2011).

Sources of chemical energy exist if reactions are not at equilibrium under environmental conditions. The degree of disequilibrium can be quantified by the chemical affinity (Anderson 2005), representing the amount of Gibbs energy ($G$) that would be released per mole of reaction progress ($\xi$) at constant temperature and pressure. Affinity of reaction $r$ is given by

$$A_r = -\left(\frac{\partial G}{\partial \xi_r}\right)_{P,T}. \tag{5}$$

A positive affinity means that the forward reaction is thermodynamically favored, while the reverse reaction is favored if the affinity is negative. Zero affinity corresponds to chemical equilibrium, where no free energy can be extracted from the reaction in either direction.

I seek to evaluate the thermodynamic potential for two oxidation-reduction (redox) reactions to serve as sources of chemical energy on K2-18b. Both reactions involve methane formation from more oxidized forms of carbon. The first reaction is the classic methanogenesis reaction starting from $CO_2$ and $H_2$ (Thauer et al. 2008), as shown below

$$CO_2(g) + 4H_2(g) \rightarrow CH_4(g) + 2H_2O(liq). \tag{6}$$

By adopting standard states of ideal gases at 1 bar and temperature $T$, and pure liquid water at pressure $P$ and temperature $T$ (these define the equilibrium constant, $K$), we can relate the reaction affinity to gas fugacities at the surface of the ocean (while relying upon the usual approximation that the activity of $H_2O$ in the liquid phase is unity; e.g., Robinson 1954). Thus, it can be shown that the relationship for $CO_2$ hydrogenation is

$$A_{CO_2} \approx 2.3026 RT \left( \log K_{CO_2} + \log f_{CO_2} + 4 \log f_{H_2} - \log f_{CH_4} \right), \qquad (7)$$

where $R$ = 8.3145 J mol$^{-1}$ K$^{-1}$ and $T$ is in Kelvin. The fugacities needed to solve this equation are obtained following the approach described in Section 2.1.

The second potential metabolic reaction is the hydrogenation of dimethyl sulfide (DMS; also called 2-thiapropane). Madhusudhan et al. (2023a) suggested that there seem to be potential signs of DMS in the atmosphere of K2-18b, although they cautioned that the identification is of low statistical significance and yet to be confirmed robustly. Nevertheless, it is worth exploring what a detection would mean since it could be confirmed later. Moreover, DMS can serve as an example to explore more generally if partially oxidized sources of organic carbon could provide energy to life on hycean worlds. In this case, the relevant reaction is

$(CH_3)_2S(g) + 2H_2(g) \rightarrow 2CH_4(g) + H_2S(g),$ \hfill (8)

and the corresponding affinity expression is

$$A_{DMS} = 2.3026 RT \left( \log K_{DMS} + \log f_{DMS} + 2 \log f_{H_2} - 2 \log f_{CH_4} - \log f_{H_2S} \right). \qquad (9)$$

$H_2S$ and DMS fugacities are estimated by performing non-ideal gas calculations using REFPROP, with the $H_2$-He-$CH_4$-$CO_2$-$H_2O$ composition from Section 2.1 and assumed mole fractions of $H_2S$ and DMS in dry air. It is assumed that $H_2S$ and DMS are present in trace quantities so that equation 1 is still satisfied. Because DMS is not in the current version of REFPROP (Huber et al. 2022), I assume that propane can serve as a suitable proxy. Out of the available options in REFPROP, dimethyl ether (DME) is chemically most similar to DMS, but REFPROP returns an error stating that it is unable to estimate the necessary mixing parameters with DME. Propane appears to be the next most similar analogue, in which a $CH_2$ unit replaces the sulfur atom in DMS (see Appendix B).

### 2.2.2. Amino Acid Synthesis

Owing to their vital role in Earth life, amino acids have figured prominently in plans to search for biosignatures elsewhere in the solar system (e.g., MacKenzie et al. 2022). It is probably impossible to find amino acids on exoplanets via telescopic observations, yet we can still assess whether exoplanetary environmental conditions might be conductive to amino acid synthesis.

Following the geochemical literature (Amend & Shock 1998; Shock & Canovas 2010), I ask, would there be affinity to form amino acids if K2-18b is a hycean world? I consider the two simplest amino acids: glycine and alanine. Both are used by Earth life but can also be produced abiotically (e.g., Miller 1953; Kvenvolden et al. 1970). Glycine can be considered a conservative case as it is among the most oxidized of the 20 protein-forming amino acids. The synthesis of glycine and alanine can be represented by the following net reactions

$2CO_2(g) + NH_3(g) + 3H_2(g) \rightarrow C_2H_5NO_2(glycine,aq) + 2H_2O(liq)$ (10)

and

$3CO_2(g) + NH_3(g) + 6H_2(g) \rightarrow C_3H_7NO_2(alanine,aq) + 4H_2O(liq)$. (11)

Chemical affinities are given by

$$A_{Gly} \approx 2.3026RT \left( \log K_{Gly} + 2\log f_{CO_2} + \log f_{NH_3} + 3\log f_{H_2} - \log a_{Gly} \right)$$ (12)

and

$$A_{Ala} \approx 2.3026RT \left( \log K_{Ala} + 3\log f_{CO_2} + \log f_{NH_3} + 6\log f_{H_2} - \log a_{Ala} \right),$$ (13)

respectively. The standard state of aqueous amino acids is an ideal 1 molal (mol of solute per kg of H$_2$O) solution referenced to infinite dilution at $T$ and $P$. I set amino acid activities ($a$) to a value of 10$^{-6}$, which is what Shock & Canovas (2010) also considered to be meaningful. This is approximately equivalent to micromolal concentrations of dissolved amino acids.

Equilibrium constants were obtained from the CHNOSZ package (Dick 2019) based on thermodynamic data from Kelley (1960), Wagman et al. (1982), Johnson et al. (1992), Richard (2001), Dick et al. (2006), and Kitadai (2014). Log $K$ values can be reproduced using parameters given in Table A2.

### 3. Results and Discussion

*3.1. Chemical Affinities for Methanogenesis*

3.1.1. From Carbon Dioxide

The feasibility of CO$_2$-reducing methanogenesis to serve as an energy source for life can be assessed by calculating the chemical affinity of this reaction as a function of environmental conditions.

Two major unknowns on K2-18b are the pressure and temperature at the surface of an ocean. Madhusudhan et al. (2023b) suggested that hycean worlds (of which K2-18b is a candidate) could have surface pressures between 1 and 1000 bar; this is the range shown in Figure 1. To avoid an extreme greenhouse effect (Innes et al. 2023; Pierrehumbert 2023), lower pressures (e.g., on the order of $10^1$ bar or less) would be more consistent with habitability on K2-18b. I consider temperatures of 25 and 120°C. The former is the thermochemical reference temperature and is similar to Earth's average surface temperature of ~15°C (Lodders & Fegley 1998). Surface temperatures below ~25°C may be implausible on K2-18b because they would promote the formation of clathrate hydrates (e.g., Marounina & Rogers 2020), which could remove $CH_4$ and $CO_2$ from the atmosphere. Such sequestration may be incompatible with the observed abundances of $CH_4$ and $CO_2$ (Madhusudhan et al. 2023a). The temperature "upper limit" in Figure 1 is not a true upper limit defined by observational data. That remains unknown. I selected a temperature of 120°C as it corresponds closely to the currently known upper bound for microbial growth on Earth (122°C; Takai et al. 2008). Because the present study is focused on habitability, this seems like a useful upper bound to consider. An important corollary, however, is that my results will depend on K2-18b having an ocean at habitable temperatures. This is a topic of debate (see Section 1). Nevertheless, if such a situation exists, we should be prepared to understand the implications for metabolic processes. This is the purpose of our exploration of these temperature-pressure conditions.

**Figure 1.** Amount of Gibbs energy that could be harnessed by synthesizing $CH_4$ from $CO_2$ and $H_2$ at the surface of an ocean on exoplanet K2-18b. The unknown atmospheric pressure is a key driver on hycean worlds. Solid vs. dashed curves show how changing the temperature from 25 to 120°C also affects energy availability. Dashed curves start at 3 bar to make sure the system is on the liquid side of the saturation curve of water. A nominal case with 1% mixing ratios of $CO_2$ and $CH_4$ is indicated by the black curve. Colored curves represent cases where the abundances of $CO_2$ and $CH_4$ are at their 1σ limits in ways that maximize (high $CO_2$-low $CH_4$) or minimize (low $CO_2$-high $CH_4$) the affinity. Equilibrium corresponds to zero affinity. The estimated affinity range for Enceladus's subsurface ocean (Waite et al. 2017) and a canonical value for the minimum affinity needed to support methanogenic life on Earth (Hoehler 2004) are shown for comparison.

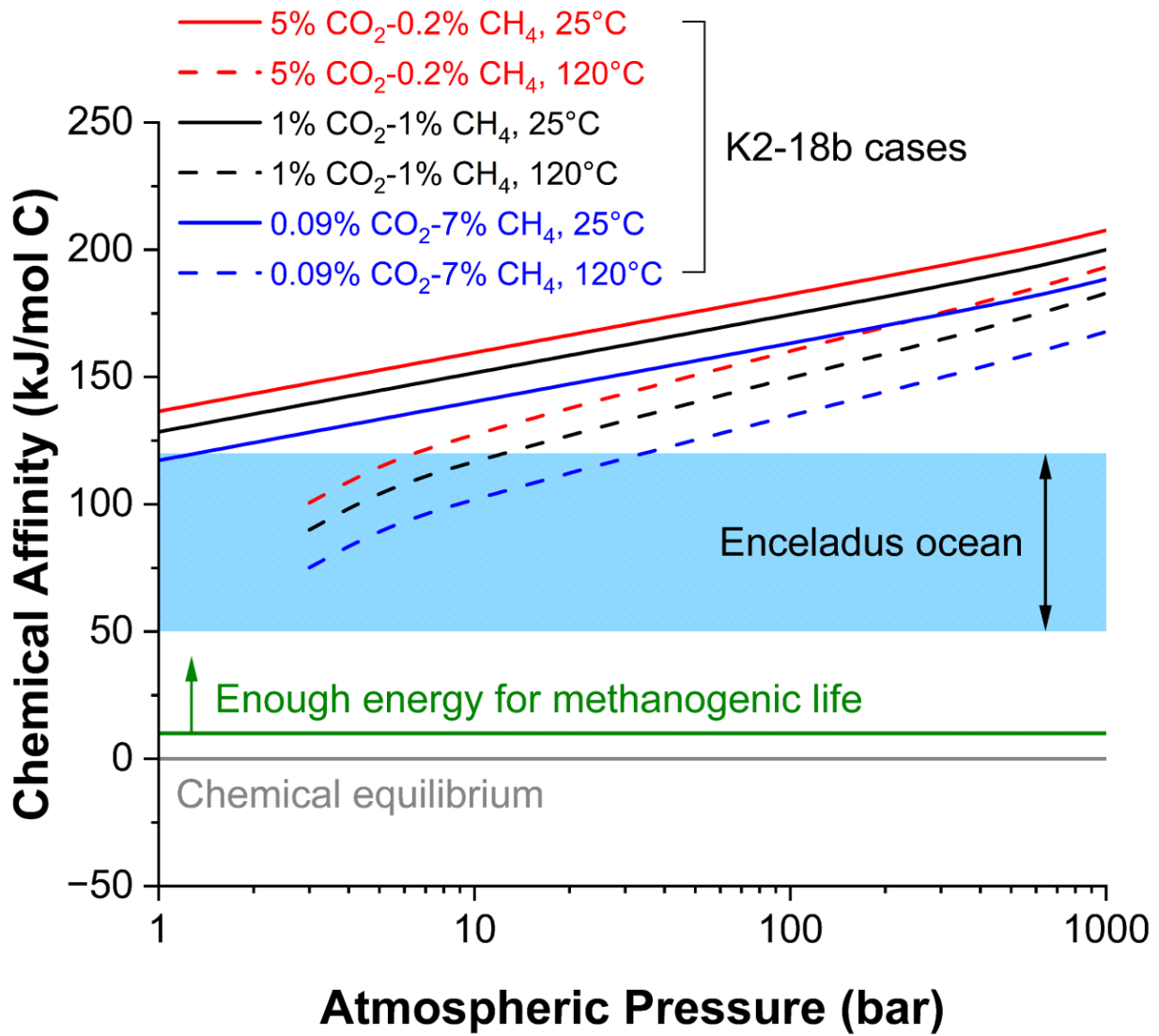

Figure 1 shows that if K2-18b has an ocean at habitable temperatures, then the observed atmospheric composition implies a large degree of chemical disequilibrium. Here, I consider compositions bounded by the 1σ ranges reported by Madhusudhan et al. (2023a). The $CO_2$-rich endmember has 5% $CO_2$ and 0.17% $CH_4$, while the $CO_2$-poor endmember has 0.093% $CO_2$ and 7.1% $CH_4$ (dry mixing ratios). The rest of the gas phase is composed of $H_2$, He, and $H_2O$. It should be noted that the use of fugacities does not mean that metabolism would occur in the gas phase. Fugacities provide a valid description as long as the atmosphere is in equilibrium with the ocean. It might also help to clarify that the chemical affinity is equal to the negative of the Gibbs energy change of a reaction under non-standard conditions (Anderson 2005).

I find that systems with more $CO_2$ and $H_2$ (reflected by the total pressure) create a stronger thermodynamic drive for autotrophic methanogenesis (Figure 1), consistent with Le Chatelier's principle. It can be seen that temperature has an inverse effect – the affinity of methanogenesis is more positive at lower temperature. This effect originates from the reaction being exothermic (heat-releasing). It is well-known in geochemistry that lower temperatures favor many hydrogenation reactions (Shock et al. 2000). Despite some variability, the calculated affinity is well above zero over the entire parameter space of interest, with values of 75-208 kJ (mol C)$^{-1}$. This implies that a biotic origin of methane on K2-18b is possible (Madhusudhan et al. 2023a; Wogan et al. 2024).

Additional insights can be obtained by making comparisons to other affinity data. Perhaps the most interesting point of comparison is the minimum Gibbs energy needed to sustain methanogens. This value is thought to be ~10 kJ (mol C)$^{-1}$ (Hoehler 2004). It is evident from examining Figure 1 that the curves for K2-18b are significantly above this threshold. Therefore, K2-18b would be energetically habitable if it has an ocean at temperatures below 120°C. This could be the first hint of a chemical energy source on a potential ocean-bearing planet beyond the solar system. Evidence of this energy source was also found at Saturn's moon Enceladus (Waite et al. 2017). That finding attracted great interest, yet an ocean on K2-18b could have a comparable or greater thermodynamic potential to support $CO_2$-reducing methanogens (Figure 1). The relationship between them depends on K2-18b's temperature-pressure conditions. For both bodies, the underlying reason why methane formation from $CO_2$ and $H_2$ is so exergonic (Gibbs energy-releasing) is because both reactants are abundant and temperatures are low (or suggested to be relatively low in the case of K2-18b; Madhusudhan et al. 2023a). If true, the combination of enormous amounts of $H_2$ and a cool/warm ocean may make K2-18b (and other hycean worlds) more energetically habitable to methanogenic life than Earth's hydrothermal systems ($A_{CO_2}$ < 130 kJ (mol C)$^{-1}$; Shock & Canovas 2010).

### 3.1.2. From Dimethyl Sulfide

We can determine whether methanogenesis from DMS would yield Gibbs energy in an ocean on K2-18b, assuming that temperatures are habitable (see Section 3.1.1). Results of this evaluation are shown in Table 1. Unlike methanogenesis from $CO_2$, the DMS reaction exhibits negligible dependence on atmospheric pressure. This is because the number of gas molecules is the same on both sides of equation 8. With no change in the number of gas molecules, the dependence on total pressure cancels out except for its effects on the fugacity coefficients. REFPROP calculations revealed that the fugacity coefficient quotient (products over reactants) for equation 8 nearly cancels out as well (to within ~30%). DMS disequilibrium depends on the fugacities of DMS and $H_2S$ at the surface of the ocean, which are proportional to their mixing ratios in the atmosphere. At present, these are poorly constrained. From their atmospheric retrievals, Madhusudhan et al. (2023a) gave a range of values for the mixing ratio of DMS from $10^{-10.1}$ to $10^{-3.7}$ at 1σ. This is the dry mixing ratio since JWST did not detect water vapor. The formal upper bound seems too high as it would imply production of DMS, probably biological in nature, at incredibly rapid rates (Madhusudhan et al. 2023a). Thus, Table 1 displays values of $10^{-10}$ and $10^{-6}$ for DMS. Because the abundance of $H_2S$ in K2-18b's atmosphere has not been reported, I consider a wide range of dry mixing ratios between $10^{-9}$ and $10^{-3}$.

**Table 1.** Chemical affinities for methane formation from dimethyl sulfide and molecular hydrogen, as a function of key environmental parameters (surface temperature, mole fractions of DMS and $H_2S$ in dry air) whose values are uncertain or unknown on K2-18b.

| $T$ (°C) | $y_{DMS,dry}$ | $y_{H_2S,dry}$ | $A_{DMS}$ (kJ/mol C) [a,b] |
|---|---|---|---|
| 25 | $10^{-10}$ | $10^{-9}$ | 79 |
| 25 | $10^{-10}$ | $10^{-6}$ | 70 |
| 25 | $10^{-10}$ | $10^{-3}$ | 62 |
| 25 | $10^{-6}$ | $10^{-9}$ | 90 |
| 25 | $10^{-6}$ | $10^{-6}$ | 82 |
| 25 | $10^{-6}$ | $10^{-3}$ | 73 |
| 120 | $10^{-10}$ | $10^{-9}$ | 83 |
| 120 | $10^{-10}$ | $10^{-6}$ | 71 |
| 120 | $10^{-10}$ | $10^{-3}$ | 60 |
| 120 | $10^{-6}$ | $10^{-9}$ | 98 |
| 120 | $10^{-6}$ | $10^{-6}$ | 87 |
| 120 | $10^{-6}$ | $10^{-3}$ | 75 |

[a] For a canonical case with 1% atmospheric methane. A one order-of-magnitude increase in the fugacity of methane would decrease the affinity by ~6-8 kJ (mol C)$^{-1}$ from 25-120°C;

[b] Variation with total pressure not shown as it is very small (<0.5 kJ (mol C)$^{-1}$).

While plausible uncertainties in the abundances of sulfur species exert some effect on energy availability (see Table 1), the dominant effect is from the high $H_2/CH_4$ ratio on K2-18b, which drives the reaction forward. The affinity for DMS hydrogenation is 62-90 or 60-98 kJ (mol C)$^{-1}$ at 25 or 120°C, respectively. These numbers are normalized per mole of carbon so that different metabolic reactions can be compared. It is striking that these affinities are all well above zero. This implies that methanogens could obtain energy by producing $CH_4$ and $H_2S$ from DMS and $H_2$. At a given temperature, the energy yield is generally lower than that from $CO_2$ hydrogenation (see Figure 1) but still significant. For comparison, Rogers & Schulte (2012) suggested that affinities between ~20 and ~43 kJ (mol C)$^{-1}$ for this reaction could support methanogens in hydrothermal ecosystems on Earth.

Three interpretations emerge from these data. First, if a non-negligible quantity of DMS is present in the atmosphere of K2-18b (Madhusudhan et al. 2023a), then a second source of chemical energy is now identified. This points to the potential for metabolic diversity if K2-18b is a hycean world. My second point is that the results for DMS could be a harbinger of the potential for other partially oxidized organics to support heterotrophic methanogenic metabolisms. DMS is not very oxidized as it has a carbon oxidation state of -2. The finding that this conservative case can provide significant energy bodes well for other, more oxidized organic compounds (e.g., acetate, formaldehyde) to contribute to energy supplies on K2-18b and similar sub-Neptunes. Third, this discussion underscores how DMS can play an important role in habitability on $H_2$-bearing ocean worlds. Much of the discussion on DMS thus far (Madhusudhan et al. 2023a; Wright 2023) has focused on whether it can be interpreted as a product of life (i.e., a biosignature). Here, I suggest that it may also be useful to consider DMS as food for life (i.e., a habitability indicator), even if DMS might have a novel abiotic origin.

### 3.2. Chemical Affinities to Synthesize Amino Acids

I now test whether amino acid synthesis would be thermodynamically favored under hycean conditions on K2-18b (Madhusudhan et al. 2023a; 2023b). This can be addressed in a similar manner as in Section 3.1.1. A notable difference, however, is that the abundance of $NH_3$ gas must be specified. We do not know the abundance of $NH_3$ on K2-18b; JWST only provided a 2σ upper limit on the dry mixing ratio (<$10^{-4.46}$; Madhusudhan et al. 2023a). Still, this is a constraint. I also consider a value more than ~30,000 times smaller, or $y_{NH_3,dry}$ = $10^{-9}$. While not a strict lower limit, this second value is more conservative and can help us understand the sensitivity of synthesis affinities to the $NH_3$ abundance. An intermediate mixing ratio of $10^{-6}$ is considered as well.

Figure 2 shows calculated affinities for the formation of glycine and alanine from $CO_2$, $NH_3$, and $H_2$ at temperatures of 25 and 120°C. These values correspond to a nominal atmosphere containing ~1% $CO_2$ and ~1% $CH_4$ (Madhusudhan et al. 2023a). Affinities are basically always positive if K2-18b is a cool hycean world (Figure 2). Organisms could therefore make these amino acids without a net expenditure of free energy. Instead, they would release energy that could be put to other uses (e.g., peptide bond formation; Dick & Shock 2021). This acquisition of both organics and energy has been referred to as being "given a free lunch that you are paid to eat" (Shock et al. 1998). If there is an ocean that is relatively cool, then the favorable thermodynamic potential to synthesize at least these amino acids would make it easier for chemosynthetic life to thrive. These affinities do depend on the unknown abundance of atmospheric $NH_3$, but the dependence is weak. I find that variability in the mixing ratio of $NH_3$ can shift them by ~10 kJ (mol C)$^{-1}$ (Figure 2). The dominant compositional driver of these reactions is the $H_2$ fugacity, in accordance with the law of mass action (e.g., equations 12 and 13). However, some $NH_3$ needs to be present. We should be mindful that thermodynamic evaluations are agnostic on how mechanistically difficult it might be to synthesize amino acids.

**Figure 2.** Amount of Gibbs energy it costs (negative affinities) or can pay (positive affinities) to synthesize the amino acids (a) glycine and (b) alanine from $CO_2$, $NH_3$, and $H_2$ at the surface of an ocean on exoplanet K2-18b. The unknown atmospheric pressure is a key driver on hycean worlds. Solid vs. dashed curves show how changing the temperature from 25 to 120°C also affects the energetic state of these reactions. Dashed curves start at 3 bar to make sure the system is on the liquid side of the saturation curve of water. Colored curves represent cases with the indicated dry mixing ratio of $NH_3$. Equilibrium corresponds to zero affinity. Affinities in seawater and mixed fluids at the Lost City Hydrothermal Field are shown for comparison. Note the ordinate axis has a break in both plots.

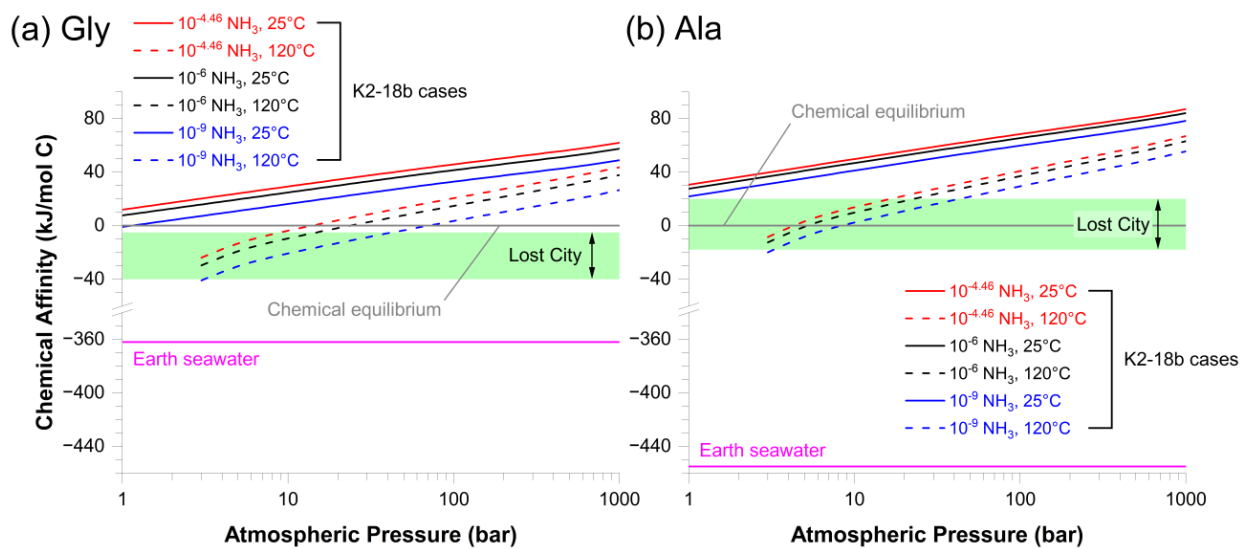

At warmer conditions, affinities for amino acid synthesis are lower (Figure 2). It then becomes more important to obtain constraints on the atmospheric pressure. It can be deduced that the pressure would need to be >20 bar for glycine synthesis to be thermodynamically favorable at 120°C, while alanine synthesis at 120°C can have positive affinities above ~4 bar. The main reason why alanine is favored over glycine is because the average oxidation state of carbon is lower in alanine (0) than in glycine (+1). Thus, $H_2$-rich hycean conditions tend to make alanine more stable than glycine. However, in either case, the atmosphere does not need to be very thick to create positive-affinity states.

Alternatively, even when affinities are negative, they have small values (Figure 2). Hence, amino acid synthesis would not cost much. An ocean on K2-18b could be thought of as analogous to a warm hydrothermal fluid. The Lost City Hydrothermal Field, located in the mid-Atlantic Ocean (Kelley et al. 2001), makes a useful point of comparison. Hydrothermal fluids there are warm and rich in $H_2$ (Seyfried et al. 2015). To constrain synthesis affinities for glycine and alanine in fluid mixtures at Lost City, I performed metastable speciation calculations (see Appendix C) with published analytical data (Table C1).

The results over a range of 25-116°C are represented by the green bands in Figure 2. It is evident that the present K2-18b models imply affinities that are comparable or more favorable for amino acid synthesis than at Lost City, which is known to support microbial life (Schrenk et al. 2004). When the affinity is negative, microbes can synthesize their building blocks by coupling the relevant reaction to exergonic reactions (see Russell & Cook 1995). The geochemical environment on K2-18b may be conducive to such coupling via energy-releasing methanogenesis reactions (e.g., Figure 1).

The most familiar comparison is also the starkest – between K2-18b and modern seawater (Figure 2). For the sake of internal consistency, chemical affinities of a seawater sample (see Table C1) were computed following the same methodology from Section 2.2.2. Affinities in these two environments are vastly different because of their huge differences in reduction potential. Glycine and alanine are much more stable in hycean systems than in $O_2$-bearing seawater. This echoes the argument that our everyday experience of large energy expenditures being needed to synthesize organic compounds does not apply to reduced environments (Shock 1990; McCollom & Amend 2005). The finding that glycine, a relatively oxidized amino acid, could have favorable affinities on this exoplanet is encouraging for the thermodynamic feasibility to synthesize other biologically relevant compounds (Amend et al. 2013). This is not to say that atmospheric chemistry cannot also produce key organic building blocks (Madhusudhan et al. 2023b). The occurrence of hycean conditions would therefore have implications to the possible origin of life (Martin et al. 2008) and how the environment may influence the composition of life (Dick et al. 2023), as yet unknown on K2-18b.

## 4. Concluding Remarks

This paper took us down a new path of exploring exoplanetary habitability. I adopted the interpretation that the sub-Neptune exoplanet K2-18b is a hycean world (Madhusudhan et al. 2023a). This planet could have a liquid water ocean beneath a $H_2$-rich atmosphere, and the ocean could be at temperatures where microbial life can survive and proliferate. Beyond the suggested presence of liquid water, it is worth seeing where this interpretation may lead in terms of the possibilities for habitability in unexpected places. I developed a thermodynamic framework that demonstrates how constraints from JWST on the atmospheric composition enable a fundamentally new type of exoplanetary study. Observations can now feed into models of the bioenergetic foundations of possible chemosynthetic biospheres on exoplanets. This approach resonates strongly with recent research on ocean worlds in our solar system (Waite et al. 2017).

I computed how far a set of metabolic reactions would be from equilibrium under assumed hycean conditions (25-120°C, 1-1000 bar). I found that methane formation from compounds that have been confidently ($CO_2$) or tentatively (DMS) identified in K2-18b's atmosphere can release substantial amounts of free energy (Figure 1; Table 1), which could be captured to power microbial metabolism. There is more than enough chemical energy to support methanogenic organisms analogous to those found on Earth. Thermodynamic calculations revealed that the synthesis of the simplest amino acids (glycine, alanine) can also be supported by chemical disequilibria in the environment, even when as little as ~1 ppbv of $NH_3$ is present in the atmosphere (Figure 2). Because of the relatively low temperature, high fugacity of $H_2$, and availability of $CO_2$, organic compounds are expected to be unusually stable in a hycean environment, much like in warm fluids surrounding some hydrothermal vents on Earth. If K2-18b is a hycean world, then these findings would provide evidence that this planet not only has liquid water but would also be able to satisfy the energy requirement of life (on both the substellar and dark sides of the planet), and may host geochemical conditions that favor the formation of organic molecules relevant to life as we know it. Overall, there seems to be an interesting possibility for habitability on K2-18b. However, I cannot emphasize enough that the present model results should not be mistaken as evidence of biological activity.

Much more work needs to be done on K2-18b. Now that excitement has built up, it can be anticipated that alternative models will emerge to challenge the hycean interpretation (e.g., Shorttle et al. 2024; Wogan et al. 2024). A next step will be to decide if the hycean interpretation is reasonable, or if K2-18b might be a hycean world in disguise. Ideally, different models will make predictions that can be discriminated by the next JWST observations of K2-18b in 2024 and 2025 (JWST C1 GO 2372, JWST C1 GO 2722). Molecular ratios should be identified that can serve as thermometers and barometers to constrain temperatures and pressures deeper in the atmosphere or at the surface of the planet (e.g., Yu et al. 2021). CO may be a robust tracer of hot, reducing conditions. DMS and $NH_3$ should also continue to be searched for, and $H_2S$ is another species that should be included in atmospheric retrievals of K2-18b. If something can't be found, an upper limit on its abundance can often be illuminating. Also, future bioenergetic modeling could explore a richer variety of organic synthesis reactions under an expanded set of geochemical conditions.

Irrespective of K2-18b's status as a hycean world (see Section 1), it is important to understand if this class of planets exists and how habitable they might be. The first issue is largely within the domain of observational astronomy. The present study can aid the latter effort by providing a template to connect atmospheric composition measurements to bioenergetic potential. Another contribution from this work is the recognition of the importance of abiotic oxidation processes in generating chemical disequilibria. Since $H_2$ would be abundant on hycean worlds, we need a better understanding of oxidant budgets. For example, how are $CO_2$ and DMS, perhaps, produced on K2-18b? While this study was limited to a static view of habitability, one can imagine that there would need to be a mechanism to recycle $CO_2$ if the planet were inhabited for a large part of its ~2.4 Gyr history (Guinan & Engle 2019). Otherwise, microbes might run out of $CO_2$ after making methane and organic compounds (see equations 6, 10, and 11). Gaidos et al. (1999) described this state of affairs as "thermodynamics-driven extinction" in the context of an ice-covered ocean on Jupiter's moon Europa. However, an ocean on K2-18b would likely be coupled to much stronger drivers of oxidation than on Europa. These include the potent influence of stellar irradiation and potentially higher-temperature hydrothermal activity. Clarifying the roles of atmospheric chemistry (promoted by lightning and high fluxes of energetic photons; dos Santos et al. 2020) and interior processes (e.g., Kite & Ford 2018; Wogan et al. 2020; Hu et al. 2021), and how they may feed into dynamic habitability, will go a long way toward generalizing our snapshot of K2-18b to other possible hycean worlds (Stevenson 1999; Pierrehumbert & Gaidos 2011; Mol Lous et al. 2022). This may, in turn, yield fresh insights into the ways that planets can support life.

## Acknowledgments

This work was supported by Heising-Simons grant 2023-4657, Simons Collaboration on the Origins of Life grant 511570FY20, NASA grant NNN13D485T, and internal funding from SwRI. It would not have been possible without the efforts of many who made JWST the great observatory of our time. The comments of an anonymous reviewer helped to improve this paper. I thank Jeff Dick for guidance on using CHNOSZ, and Xinting Yu for always stimulating discussions on exoplanets. I would also like to give thanks to Chris Perry for pulling me away from the daily grind, which provided time to nucleate the ideas explored in this paper. I dedicate this work to the memory of Jan Amend, a friend who pioneered calculations of amino acid stability in hydrothermal systems.

## Appendix A

## Regression Equations for Thermodynamic Data

Table A1 shows an equation for the fugacity of liquid water, together with parameter values at temperature grid points that are commonly used in geochemical modeling (e.g., Bethke 2022). The applicable pressure range is also noted.

**Table A1.** Empirical parameters to calculate the fugacity of liquid water as a function of total pressure at different temperatures.

| Temperature (°C) | Pressure range (bar) | $\log f_{H_2O(liq)}$ (bar) = $A + B \times P$ (bar) [b] | |
|---|---|---|---|
| | | A | B |
| 0 | 1-2000 | -2.2123 | 3.32E-04 |
| 25 | 1-2000 | -1.4987 | 3.06E-04 |
| 60 | 1-2000 | -0.7017 | 2.78E-04 |
| 100 | 1-2000 | 0.0001 | 2.54E-04 |
| 120 | 2-2000 [a] | 0.2884 | 2.44E-04 |
| 150 | 4.8-2000 [a] | 0.66 | 2.32E-04 |

[a] The total pressure must be higher than the saturation vapor pressure of liquid water for the liquid phase to be stable;

[b] This equation reproduces numerical values of fugacity from REFPROP to within ~1%.

An equation for the common logarithm of the equilibrium constant of methanogenesis and amino acid synthesis reactions is shown in Table A2. The parameters given in the table account for the effects of temperature and pressure.

**Table A2.** Regressed parameters that can be used to compute equilibrium constants of metabolic reactions as a function of temperature (0-150°C) and total pressure (1-2000 bar).

| Reaction | $\log K = A + (B + CP) \times T^{-1}$ [e, f] | | |
|---|---|---|---|
| | A | B | C |
| $CO_2(g) + 4H_2(g) \rightarrow CH_4(g) + 2H_2O(liq)$ [a] | -21.249 | 13167.6 | -0.1858 |
| $(CH_3)_2S(g) + 2H_2(g) \rightarrow 2CH_4(g) + H_2S(g)$ [b] | 1.443 | 6990 | 0 [g] |
| $2CO_2(g) + NH_3(g) + 3H_2(g) \rightarrow C_2H_5NO_2(aq) + 2H_2O(liq)$ [a, c] | -37.224 | 13190.4 | -0.4206 |
| $3CO_2(g) + NH_3(g) + 6H_2(g) \rightarrow C_3H_7NO_2(aq) + 4H_2O(liq)$ [a, d] | -60.559 | 24378.3 | -0.6958 |

[a] The total pressure must be higher than the saturation vapor pressure of liquid water for the liquid phase to be stable;

[b] Dimethyl sulfide hydrogenation;

[c] Glycine synthesis;

[d] Alanine synthesis;

[e] Pressure in bar and temperature in K;

[f] This equation reproduces log $K$ values from CHNOSZ to within ~0.1 log units;

[g] No pressure dependence.

**Appendix B**

**Strategy for Dealing with Poorly Characterized Gas Mixtures**

REFPROP calculates the thermodynamic properties of fluid mixtures (including gases) from the properties of pure fluids using binary mixing parameters. These parameters were determined by fitting a mixing model to experimental data (e.g., speed of sound, second virial coefficient). While REFPROP includes a comprehensive suite of binary parameters (see Kunz & Wagner 2012), I noticed that some are missing or uniformly set to a value of 1, indicating that the binary has not been regressed. The binaries that need to be addressed for the model of K2-18b's atmosphere are: $H_2$-$H_2O$ and He-$H_2O$ (for the major species composition); $H_2$-$NH_3$, He-$NH_3$, and $CO_2$-$NH_3$ (for N-bearing systems); and $H_2$-$H_2S$, He-$H_2S$, and He-$C_3H_8$ (for S-bearing systems in which propane, $CH_3$-$CH_2$-$CH_3$, is assumed to approximate the behavior of DMS, $CH_3$-S-$CH_3$).

I developed a strategy to address each of these issues while not introducing meaningful inaccuracies. For the baseline system of $H_2$-He-$CH_4$-$CO_2$-$H_2O$, I retained the default REFPROP model that assigns 1s to all parameters for the $H_2$-$H_2O$ and He-$H_2O$ binaries. I suspect that this will lead to inaccuracies in how the model treats $H_2$-$H_2O$ and He-$H_2O$ interactions, but these inaccuracies will not be significant. At high pressures where non-ideal corrections are needed most, the gas mixture is practically devoid of water because of the low (<4 bar) saturation fugacity of liquid water at the temperatures of interest in this study. At low pressures, water can be a major species in the gas phase, especially at 120°C. Nevertheless, the magnitude of the non-ideal correction factors (i.e., the fugacity coefficients) is nearly unity at low pressures. The gas is essentially ideal, so it does not matter whether intermolecular interactions are modeled accurately. I performed a series of calculations to verify these expectations. I found that the behavior of the mixture does converge to that of the $H_2$-He-$CH_4$-$CO_2$ endmember as pressure increases. At low pressures, I replaced $H_2$+He with $N_2$ (as a proxy with well-defined binary parameters with the other components) and found that computed fugacities can differ from those from the default model by only a few percent. From the results of these tests, I conclude that the model describes the behavior of the major species sufficiently well.

For mixtures containing $NH_3$, I adopt fugacities for all major species from the baseline system since I am considering dry mixing ratios of $NH_3$ less than 0.01% (i.e., too small to perturb the behavior of the major species). I then renormalize the composition in terms of the assumed abundance of $NH_3$ and replace $H_2$+He+$CO_2$ with $N_2$, in line with a preference for proxies when data are unavailable. In test calculations with $NH_3$-free gas with all other variables held constant, I found that the fugacity coefficients of $H_2$, He, and $CO_2$ are similar to those of the equivalent amount of $N_2$ (to within ~15%). This supports the idea that it is reasonable to consider $N_2$ as a proxy for these gases under the modeled conditions. After making this replacement, I used REFPROP to estimate fugacities of $NH_3$ in atmospheric mixtures on K2-18b.

I employed an analogous approach to estimate fugacities of $H_2S$ and $DMS/C_3H_8$ in S-bearing systems. Although, here, $N_2$ replaces $H_2$+He (note that REFPROP has binary parameters for $CO_2$ with $H_2S$ and $C_3H_8$). It bears repeating that the proxy-based composition is only used to calculate fugacities for minor/trace species – $H_2S$ and $DMS/C_3H_8$ in the present case. Major species fugacities come from the first set of calculations described above. In addition, I tested how sensitive chemical affinities for equation 8 are to the use of propane as a proxy for the non-ideal gas behavior of DMS. If $H_2S$ is taken as a proxy instead, the affinities would increase (methanogenesis more favorable), but the magnitude of this shift is insignificant (<0.5 kJ (mol C)$^{-1}$; cf. Table 1) over the range of conditions considered in this work. Therefore, it does not matter which proxy is selected.

**Appendix C**

**Calculating the Energetic Costs and Benefits of Synthesizing Amino Acids in Mixed Fluids at the Lost City Hydrothermal Field**

At Lost City, warm hydrothermal fluids mix with cold seawater, producing a continuum of fluid compositions depending on distance from hydrothermal vents. To calculate affinities of amino acid synthesis reactions, fugacities of $CO_2$, $NH_3$, and $H_2$ are needed (see equations 12 and 13). I perform reaction path calculations using the React app in The Geochemist's Workbench (GWB) 2023 (Bethke 2022) to simulate the mixing process. This requires compositions and temperatures of the endmember fluids (see Table C1). Element/species molal (mol (kg $H_2O$)$^{-1}$) concentrations and in-situ pH values were derived from published data (with pH measured at 25°C) by performing speciation calculations using GWB's SpecE8 app (Bethke 2022).

**Table C1.** Geochemical properties of endmember fluids needed to perform calculations of fluid mixing at the Lost City Hydrothermal Field.

| Quantity | Vent fluid | Local seawater |
|---|---|---|
| $T$ (°C) | 116 [a] | 8 [e] |
| pH (at $T$) | 8.41 [a,b] | 8.22 [a,b] |
| Na (mmolal) | 500 [a] | 493 [a] |
| K (mmolal) | 10.8 [a] | 10.7 [a] |
| $\sum$NH$_3$ (mmolal) | 0.0036 [c] | 0.0001 [c] |
| Mg (mmolal) | ~0 [a] | 55.3 [a] |
| Ca (mmolal) | 28.3 [a] | 10.8 [a] |
| Sr (mmolal) | 0.105 [a] | 0.095 [a] |
| Cl (mmolal) | 559 [a] | 575 [a] |
| Br (mmolal) | 0.879 [a] | 0.889 [a] |
| SO$_4$ (mmolal) | 3.3 [a] | 29.2 [a] |
| $\sum$CO$_2$ (mmolal) | 0.013 [d] | 2.1 [f] |
| B (mmolal) | 0.039 [a] | 0.43 [a] |
| Si (mmolal) | 0.075 [a] | 0.026 [a] |
| H$_2$ (mmolal) | 11.2 [a] | … |
| O$_2$ (mmolal) | … | 0.18 [e] |
| $f_{CO_2}$ (bar) | 10$^{-5.4}$ [b] | 10$^{-3.4}$ [b] |
| $f_{NH_3}$ (bar) | 10$^{-5.9}$ [b] | 10$^{-10.9}$ [b] |
| $f_{H_2}$ (bar) | 10$^{1.2}$ [b] | 10$^{-44.1}$ [b] |

[a] Seyfried et al. (2015);

[b] Calculated by respeciating the published composition at the in-situ temperature;

[c] Reeves et al. (2014);

[d] Dissolved inorganic carbon content estimated by assuming that the hydrothermal fluid was in equilibrium with calcite below the seafloor;

[e] Schlitzer (2000);

[f] Estimated by assuming that the ratio of dissolved inorganic carbon to chloride is similar to that in standard seawater (Millero et al. 2008).

One complication to note is that there is uncertainty concerning the concentration of dissolved inorganic carbon (DIC) in the Lost City endmember. Reported values span a wide range of 0.0001-0.2 mmol kg$^{-1}$ (Lang et al. 2010; Reeves et al. 2014). It is difficult to determine the endmember DIC concentration because even a small amount of seawater (high DIC) contamination of vent fluid samples (low DIC) introduces substantial uncertainties. I decided to address this issue by assuming that the high Ca content of the hydrothermal endmember (see Table C1) constrains its DIC content via equilibrium with calcite ($CaCO_3$) prior to the emergence of the fluid from the seafloor. Strictly speaking, this provides an upper limit on DIC concentration since the fluid could be undersaturated with respect to calcite. Nevertheless, the calcite-buffered concentration is over two orders of magnitude lower than that in seawater, so it serves the purpose of illustrating the stark contrast between these fluids in terms of DIC concentrations that factor into $CO_2$ fugacities.

I perform mixing calculations following the procedure outlined in GWB's Online Academy lesson on hydrothermal fluids (https://academy.gwb.com/hydrothermal.php). I use a thermodynamic database generated with the PyGCC package (Awolayo & Tutolo 2022). The database contains equilibrium constants and activity parameters for aqueous species evaluated using standard approaches (Helgeson 1969; Shock et al. 1989; 1997). The total pressure in the database is 100 bar, similar to that at Lost City (a ~800 m water column equals ~80 bar).

Mixing calculations track the geochemical evolution of a hydrothermal fluid as increasing proportions of seawater are added to it (McCollom & Shock 1997). The model accounts for conservation of thermal energy, acid-base speciation, and ion complexation. Also, it allows precipitation of minerals (brucite, $Mg(OH)_2$; aragonite, $CaCO_3$) that are found in freshly formed chimneys at Lost City (Ludwig et al. 2006). All redox couples except $H_2$-$O_2$ are assumed to be kinetically inhibited over the mixing timescale. Charge is balanced on chloride. Among the model outputs are the fugacities of $CO_2$, $NH_3$, and $H_2$. I use these along with ~1 µmolal concentrations of glycine and alanine (Shock & Canovas 2010) to compute affinities for synthesizing these amino acids from inorganic precursors in mixed fluids at Lost City (Figure C1).

**Figure C1.** How chemical affinities of two amino acid synthesis reactions (see equations 10 and 11) evolve as a Lost City hydrothermal endmember fluid (116°C) mixes with surrounding seawater (8°C). Equilibrium corresponds to zero affinity. The reaction path calculations predict that brucite and aragonite should precipitate above ~81°C, whereas the fluid mixture is saturated in only aragonite at lower temperatures.

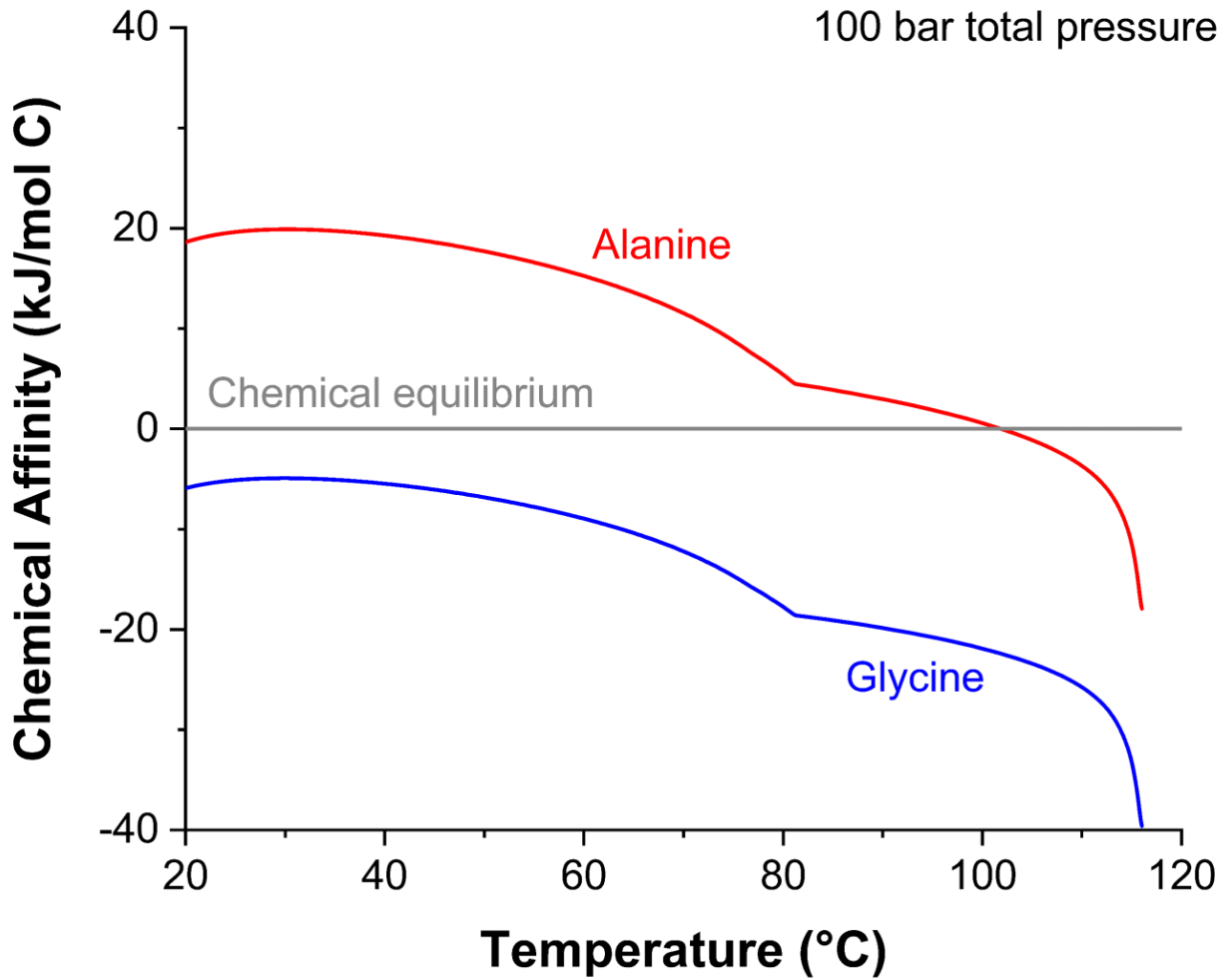